\documentclass[useAMS,usenatbib]{mn2e}
\usepackage{times}
\usepackage{graphicx}

\def\aap{A\&A}%
\def\actaa{Acta Astronomica}%
\def\aj{AJ}%
\def\apj{ApJ}%
\def\apjl{ApJ Letters}%
\def\apjs{ApJ Suppl.~Series}%

\def\mnras{MNRAS}%
\def\nat{Nature}%
\def\pasp{PASP}%
\title[Life after eruption. II.]{Life after eruption -- II. 
The eclipsing old nova V728 Scorpii}
\author[C. Tappert et al.]%
{
C. Tappert,$^{1}$\thanks{E-mail: claus.tappert@uv.cl} 
N. Vogt,$^{1}$
L. Schmidtobreick,$^{2}$ 
A. Ederoclite$^{3}$ and
J. Vanderbeke$^{2,4}$
\footnotemark[1]\thanks{Based on observations with ESO telescopes,
proposal numbers 088.D-0588 and 089.D-0505}\\
$^{1}$Dpto. de F\'{\i}sica y Astronom\'{\i}a, Universidad de Valpara\'{\i}so,
Avda. Gran Breta\~na 1111, Valpara\'{\i}so, Chile\\
$^{2}$European Southern Observatory, Alonso de Cordova 3106, Santiago, Chile\\
$^{3}$Centro de Estudios de F\'{\i}sica del Cosmos de Arag\'on, Plaza San 
Juan 1, Planta 2, Teruel, E44001, Spain\\
$^{4}$Sterrenkundig Observatorium, Universiteit Gent, Krijgslaan 281 S9, 
B-9000 Gent, Belgium
}
\begin{document}

\date{Accepted. Received}

\pagerange{\pageref{firstpage}--\pageref{lastpage}} \pubyear{2011}

\maketitle

\label{firstpage}

\begin{abstract}
The old nova V728 Sco has been recently recovered via photometric and
spectroscopic observations, 150 years after the nova eruption. The spectral
properties pointed to a high-inclination system with a comparatively low
mass-transfer rate. In this paper we show that the object is an
eclipsing system with an orbital period of 3.32 h. It has enhanced long-term
variability that can be interpreted as 'stunted' dwarf-nova-type outbursts.
Using the ingress and egress times of the eclipsed components we calculate
the radius of the central object. The latter turns out to be significantly 
larger than a white dwarf and we identify it with a hot inner disc. The
implications for models on the behaviour of post-novae are discussed. 
\end{abstract}

\begin{keywords}
binaries: close -- eclipsing -- novae, cataclysmic variables
\end{keywords}

\section{Introduction}
\defcitealias{tappertetal12-1}{Paper I}%
A nova eruption is a thermonuclear explosion on the surface of an accreting
white dwarf. In the process, a part of the outer layers of the white dwarf
is ejected as a nova shell. It is not clear if the expelled material
amounts to less or more than the accreted material, although the newest studies
point to the former possibility \citep*{zorotovicetal11-1}. In the case of 
{\sl classical novae}, the donor usually is a late-type main-sequence star.
This makes the system hosting the nova event a cataclysmic variable (CV). Since
the CV is not destroyed by the explosion, it is safe to assume that
a nova eruption is a recurrent event. A typical shell mass of 
$10^{-5}$--$10^{-4}~\mathrm{M_\odot}$ \citep[e.g.,][]{yaronetal05-1} and an 
upper limit 
of the average mass-transfer rate of $\sim10^{-8}~\mathrm{M_\odot~yr^{-1}}$ 
\citep{townsley+gaensicke09-1} then translates to recurrence times 
$>10^{3}~\mathrm{yr}$. This is now also supported observationally
\citep{sharaetal12-3}.

Nova eruptions can thus be expected to be part of the evolution of every CV, 
assuming that the 
mass-transfer rate is high enough to accumulate the necessary critical amount 
of material on the surface of the white dwarf within the lifetime of the 
system. Hence, once the effects of the nova eruption have worn off, the
behaviour of the underlying CV should be determined (once again) by the 
properties that make it part of one of the many subtypes of CVs, 
i.e.~especially its orbital period, mass-transfer rate and the magnetic field 
strength of the white dwarf. There are several post-novae that corroborate
this assumption, e.g.~DQ Her (Nova Her 1934) is the prototype intermediate 
polar \citep{warner83-1}, and RR Pic (Nova Pic 1925) belongs to the
SW Sex stars \citep*{schmidtobreicketal03-4}. Of special importance in this
context is the discovery of an ancient nova shell around the dwarf novae Z Cam
and AT Cnc \citep{sharaetal07-1,sharaetal12-4}, which made these the first CVs 
to be established as post-novae not based on the observation of the actual 
nova eruption.

Within our project to establish a sample of post-novae to compare their
characteristics as a group we have recently recovered the position of the
system V728 Sco \citep[Nova Sco 1862; ][]{tebbutt78-3} 150 yr after its
eruption \citep[][hereafter Paper I]{tappertetal12-1}. The spectral
characteristics indicated a high system inclination and a comparatively
low mass-transfer rate. Here we present a more detailed study of that object.

\section{Observations and reduction}

\begin{table}
\caption[]{Log of observations.}
\label{obslog_tab}
\begin{tabular}{lllll}
\hline\noalign{\smallskip}
Date & Filter/grism ({\AA}) & $n$ & $t_\mathrm{exp}$ (s) & $\Delta t$ (h) \\
\hline\noalign{\smallskip}
2012-03-25 & \#20 [6035--7135, 3.8]   & 3   & 900     & 1.78 \\
2012-03-26 & \#20 [6035--7135, 3.8]   & 24  & 600/300 & 2.37 \\
2012-03-27 & \#20 [6035--7135, 3.8]   & 41  & 300     & 3.84 \\
2012-03-28 & $V$                      & 299 & 20      & 4.59 \\
2012-04-01 & $V$                      & 151 & 20      & 2.54 \\
2012-05-15 & $V$                      & 61  & 40      & 1.29 \\
2012-05-16 & \#7 [3015--5190, 7.0]    & 12  & 300     & 1.04 \\
           & \#16 [5965--10250, 13.6] & 10  & 300     & 1.09 \\
\hline\noalign{\smallskip}
\end{tabular}
\end{table}

We observed V728 Sco during observing runs in 2012 March and in May,
using EFOSC2 \citep*{eckertetal89-1} mounted on the ESO-NTT at La Silla,
Chile. We employed the Bessell $V$ filter to obtain a photometric
light curve, the medium resolution grism \#20 to measure radial velocities
and the two low resolution grisms \#7 and \#16 to examine the spectral 
variation of the blue and the red part of the spectrum throughout eclipse. 
The observations are summarized in Table \ref{obslog_tab}. It states the 
start of the night of the observations and the filter or grism used for the 
photometric or spectroscopic observations, respectively. The column for the 
latter also includes the nominal wavelength range and the spectral resolution 
as the typical full width at half-maximum (FWHM) of an arc line. The actual 
useful wavelength range is somewhat shorter for the low resolution grisms: due 
to the efficiency limits the signal-to-noise ratio (S/N) decreases strongly at
wavelengths $<$3650 {\AA} (grism \#7) and $>$9500 {\AA} (grism \#16). The 
latter grism additionally suffers from fringing for wavelengths $>$7000 {\AA}. 
Table \ref{obslog_tab} furthermore contains information on the number $n$ of 
frames taken, the exposure time $t_\mathrm{exp}$ of an individual data frame, 
and the time range $\Delta t$ covered by the observations.

\subsection{Photometry}

The reduction of the photometric data includes the subtraction of bias frames, 
but no flat-field correction because EFOSC2 flats are affected by a central 
light concentration. Using {\sc iraf}'s {\sc daophot} package, relative 
photometric 
magnitudes were obtained for all stars that were found by the {\sc daofind}
algorithm. On some frames during deep eclipse, the target (V728 Sco) was not
detected by that routine, and its position was determined with respect to
a number of neighbour stars. An aperture radius of 3.0 pixels was used for
all measurements. The data subsequently served as input for the standalone
{\sc daomatch} and {\sc daomaster} routines \citep{stetson92-1}. Differential 
magnitudes for V728 Sco were computed with respect to the average of
11 suitable comparison stars in the neighbourhood ($\pm$200 pixels).
Since no standard star data were taken during these observing runs, calibrated
magnitudes were obtained by comparison with the photometric data reported
in \citetalias{tappertetal12-1}. We furthermore computed differential 
light curves for all other stars on the field, but did not find any clear 
intrinsic variability for any of them.

\subsection{Spectroscopy}

The spectroscopic data were reduced by subtracting the nightly average
bias frame and by dividing them by an average flat-field. The latter had
been normalised for its spectral energy distribution by division through
a high-order cubic spline fit along the dispersion axis. The spectra were
subsequently extracted using the implementation of the optimal extraction
algorithm \citep{horne86-1} in IRAF's {\sc apall} routine. Wavelength
calibration was performed using Thorium-Argon spectra that were taken
during the afternoon. Flux-calibrated spectra for grisms \#7 and \#16 were
obtained in comparison to the standard star LTT 7987 that was observed during
the same night. Since an absolut flux calibration was not necessary for our
purposes we did not perform any correction for slit losses. No flux standards 
were observed for the grism \#20 spectra, which nevertheless were corrected 
for the instrumental response function using previously taken data.

\section{Results}

\subsection{Light curves and the orbital period}

\begin{figure}
\includegraphics[width=\columnwidth]{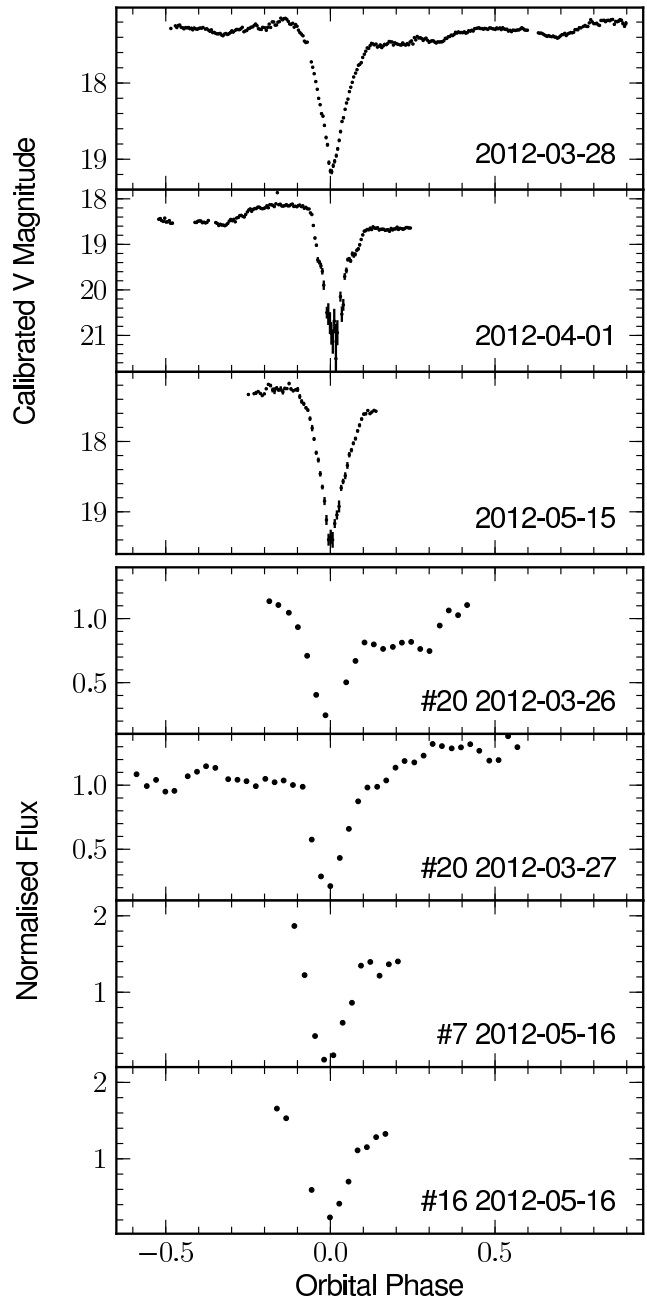}
\caption[]{Orbital light curves from the photometric $V$ data and from
normalised fluxes of the spectroscopic data taken with grisms \#20, \#7,
and \#16. The data are folded on the ephemeris given in equation 
(\ref{ephem_eq})}
\label{v728scolc_fig}
\end{figure}

\begin{table}
\caption[]{Long-term photometric data.}
\label{ltphot_tab}
\begin{tabular}{llll}
\hline\noalign{\smallskip}
HJD$-$2\,450\,000 & $R$ (mag) & $V$ (mag) & Notes\\
\hline\noalign{\smallskip}
4972.5 &            & 18.47(11) & Calibrated data \\ 
5742.5 & 17.431(03) & 18.03(01) & Acquisition frame\\ 
6012.5 & 16.41(02)  & 17.01(02) & Acquisition frame\\ 
6013.5 & 16.49(03)  & 17.08(03) & Acquisition frame\\ 
6014.5 & 16.61(02)  & 17.21(02) & Acquisition frame\\ 
6015.5 &            & 17.33(09) & Light curve\\ 
6019.5 &            & 18.43(20) & Light curve\\ 
6063.5 &            & 17.36(14) & Light curve\\ 
6064.5 & 16.81(01)  & 17.41(02) & Acquisition frame\\
\hline\noalign{\smallskip}
\end{tabular}
\end{table}

\begin{figure}
\includegraphics[width=\columnwidth]{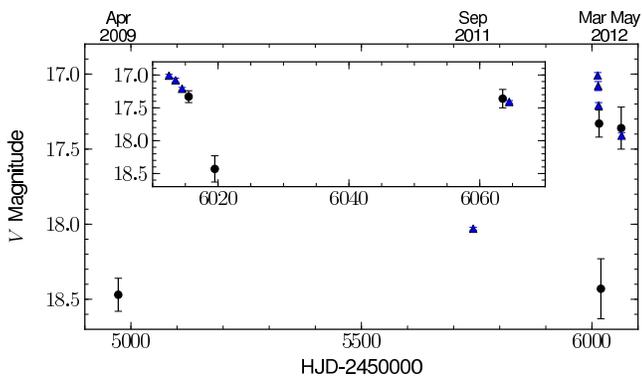}
\caption[]{Long-term light curve. The circles mark genuine $V$ photometric 
data, 
while the triangles indicate $V$ magnitudes calculated from measurements in the
$R$ band via the colour index $V\!-\!R = 0.60$ \citepalias{tappertetal12-1}.
The inset shows a close-up of the 2012 March and May data.}
\label{v728scoltlc_fig}
\end{figure}

\begin{table}
\caption[]{Eclipse timings.}
\label{ecltimes_tab}
\begin{tabular}{lllll}
\hline\noalign{\smallskip}
$t_0$        & Cycle  & Data & O$-$C & O$-$C \\ %
\hspace{0cm} (HJD$-$245\,6000) & & & ($10^{-4}$ d) & (orbits) \\
\hline\noalign{\smallskip}
13.8704(23)  & $-$14 & \#20 & 2.37     & 0.0017    \\            
14.837\,88(57) & $-$7  & \#20 & $-$6.62  & $-$0.0048 \\            
15.807\,30(23) & 0     & $V$  & 3.80     & 0.0027    \\            
19.818\,24(31) & 29    & $V$  & $-$5.34  & $-$0.0039 \\            
63.672\,88(58) & 346   & $V$  & 4.08     & 0.0029    \\            
64.6404(21)  & 353   & \#7  & $-$4.43  & $-$0.0032 \\            
64.7775(16)  & 354   & \#16 & $-$16.94 & $-$0.012  \\            
\hline\noalign{\smallskip}
\end{tabular}
\end{table}

The photometric data yield the light curves shown in the top three panels in
Fig.~\ref{v728scolc_fig}. The most prominent feature is doubtlessly the
deep eclipse. We furthermore notice the presence of an orbital hump that lets
the photometric brightness rise to a maximum just prior to the eclipse. This
hump, which most probably corresponds to the hot spot, i.e.~the region 
where the accretion from the secondary star impacts on the accretion disc, 
is more prominent in the second of the light curves, from April 
1$^\mathrm{st}$, than in the other two, from March 28 and from May 15. In 
fact, the April data show the system in an $\sim$1 mag fainter brightness 
state. This bears consequences for the determination of the orbital period. 
In CVs, a difference in the state of brightness can correspond to different 
radii of the accretion disc \citep[e.g.,][]{smak84-1}, potentially affecting 
the position and the (relative) contribution of especially the hot spot.
This, in turn, can have an effect on the phasing and the amplitudes of
radial velocity and light curves.

Table \ref{ltphot_tab} summarizes the photometric data from several
observing runs. It contains the Heliocentric Julian Day (HJD) corresponding
to the night of the observations, and photometric $R$ and $V$ magnitudes.
The first data point was taken from the calibrated multi-colour photometry
reported in \citetalias{tappertetal12-1}, which furthermore yielded a
colour index $V\!-\!R = 0.60(01)$ mag. The acquisition frame data correspond
to the spectroscopic observing runs and were taken in the $R$ passband. The
respective $V$ magnitude was calculated according to the above colour index. 
While the colour is supposed to change with the brightness of the system we
can expect that change to amount to much less than 0.5 mag \citep*[see e.g.~%
the series of papers by][]{spoglietal00-1,spoglietal00-2,spoglietal00-3}.
All data points were examined for their corresponding orbital phase, using
equation (\ref{ephem_eq}). We find that none of them is affected by the eclipse.
Specifically, the $U\!B\!V\!R$ photometry was taken during phases 0.49--0.77,
the acquisition frame of the low-resolution spectrum at phase 0.66, and
the other acquisition frames just before the respective spectroscopic time 
series shown in Fig.~\ref{v728scolc_fig}. Lastly, the light-curve data are 
represented as nightly averages of the out-of-eclipse brightness. We can thus
assume the influence of the orbital variability on the points shown
in Fig.~\ref{v728scoltlc_fig} to be minimal.

We can clearly see that V728 Sco appears to change frequently between 
brightness states. The $U\!B\!V\!R$ photometry from 2009 May 20 and the 
time series
photometry from 2012 April 1 find the system at $V \sim 18.5$ out of
eclipse, while the other light curves and the time series spectroscopic
data catch it at an $\sim 1\!-\!1.5$ mag higher brightness. Furthermore, the
low-resolution spectroscopy from 2011 June 29 apparently was taken when
the system was in between the bright and the low state. While further
implications of this behaviour are discussed in Section \ref{disc_sec},
at this point we regard the consequences with respect to the determination
of the orbital period. The light curves in Fig.~\ref{v728scolc_fig} indicate
that the contribution of the asymmetrically placed hot spot is strongest
in the low state and much fainter otherwise. We therefore assume that the
eclipsed light in high state corresponds mainly to the accretion disc.
The latter is situated symmetrically around the white dwarf, and thus
mid-eclipse should represent the superior conjunction of the primary.
In comparison, in low state the eclipse is much more structured, as different
sources -- the disc, the central object, the hot spot -- are eclipsed, and
the minimum is likely to be offset in phase with respect to superior 
conjunction. We thus determined the eclipse timings for the high-state 
data by fitting a second-order polynomial to the lowest points of the eclipses,
while mid-eclipse for the low-state data was calculated from the ingress and 
egress times of the central object (see Section \ref{ecl_sec}). These data are 
summarized in Table \ref{ecltimes_tab}. Here, the fluxes of the spectroscopic 
data were calculated as the total flux in the spectral ranges 6200--7000 {\AA}, 
4000--5000 {\AA} and 6200--8000 {\AA}, for grisms \#20, \#7 and \#16, 
respectively. They were subsequently normalized by dividing through the 
mean values. We then fitted the eclipse timings with a linear 
regression, weighting the data points according to their respective 
uncertainties. In this way we find the ephemeris
\begin{equation}
\label{ephem_eq}
T_0 ({\rm HJD}) = 245\,6015.806\,92(19) + 0.138\,340(02)~E,
\end{equation}
where $E$ is the cycle number.

We would like to point out that no two of our recorded eclipses are really 
the same. Even the two $V$-band high-state eclipses have different depth 
and show different minor asymmetries and bumps. We have tested the robustness
of our result by using different sets of eclipses for the determination
of the orbital period. All permutations yield the same result within the 
errors.

\subsection{Eclipse spectra}
\label{eclsp_sec}

\begin{figure*}
\includegraphics[width=2.0\columnwidth]{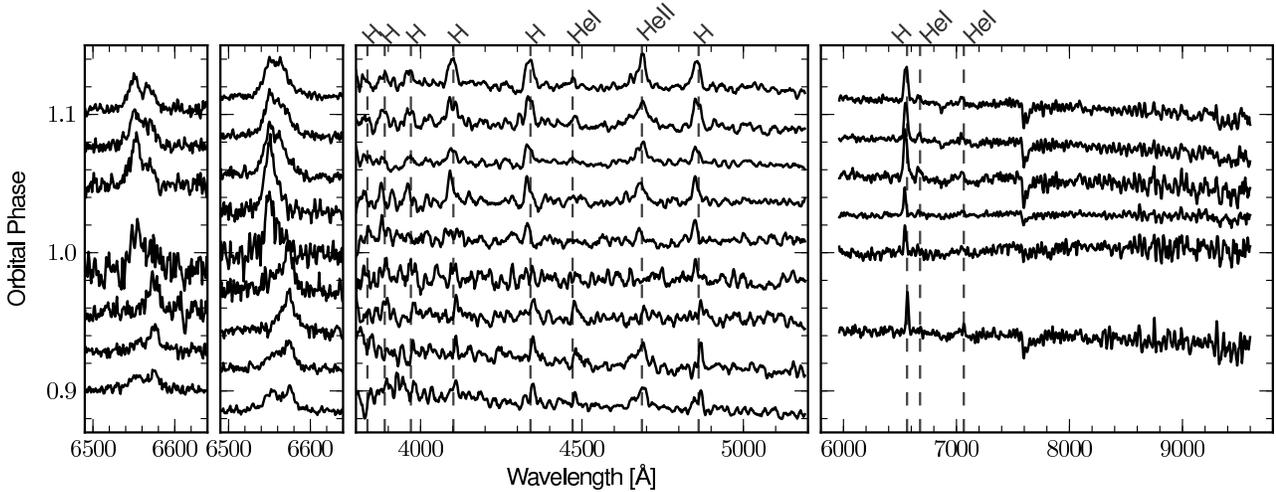}
\caption[]{Time-series spectroscopic data around the eclipse phases. From left 
to right: grism \#20 data from March 26, from March 27, grism \#7 data,
and grism \#16 data, both from May 16. The spectra have been normalised with
respect to the continuum and shifted vertically according to their 
corresponding orbital phase. The grism \#7 and \#16 spectra have been scaled 
individually to better visualise the sequence.}
\label{tracespecs_fig}
\end{figure*}

In Fig.~\ref{tracespecs_fig} we have collected the spectral time-series
throughout eclipse. These consist of the grism \#20 data that show the
behaviour of the H$\alpha$ line on two subsequent nights during decline
from an outburst, and the two low-resolution series of the blue and the
red spectral range, taken in one night with grisms \#7 and \#16, 
respectively. The long-term light curve in Fig.~\ref{v728scoltlc_fig}
suggests that also the latter data were taken during decline, at an only
slightly lower brightness state than the grism \#20 spectra. 

The high-resolution data (grism \#20, the two left-hand plots) clearly show a 
double-peaked line profile out of eclipse. Throughout eclipse (phases 
0.95--1.05) the line presents a single peak, which rapidly changes its position 
in mid-eclipse. This is the well-known ''Z wave`` behaviour, as first
the part of the accretion disc with velocities towards the observer
(that form the blue peak of the double-peak profile) and then the
part with velocities away from the observer (the red peak) are eclipsed.
This proves that V728 Sco is a disc system. 

The grism \#7 and \#16 data have too low S/N to analyse the mid-eclipse
spectra in detail, but they still present a couple of interesting
phenomena. First, it can be clearly seen that the He{\sc ii} $\lambda$4686
line that out of eclipse rivals the H$\beta$ emission line in strength
disappears completely within the limits of the S/N during mid-eclipse,
while H$\beta$ remains visible (third plot from the left). This presents
support for the existence of a hot inner disc (as discussed in Section 
\ref{ecl_sec}). Secondly, we note that the slope of the red part of the 
spectrum 
(right plot) in mid-eclipse switches from negative to positive towards longer 
wavelengths.  While such behaviour is expected during the eclipse of the hotter
parts of the system, the continuum also becomes significantly more ''bumpy`` 
indicating that at this point and wavelength range the secondary star is the 
dominant light source. For that reason, there is a good chance that a high-S/N 
spectrum taken in mid-eclipse would be able to reveal the spectral type of that
component.

\subsection{Radial velocities and Doppler tomography}

\begin{figure}
\includegraphics[width=\columnwidth]{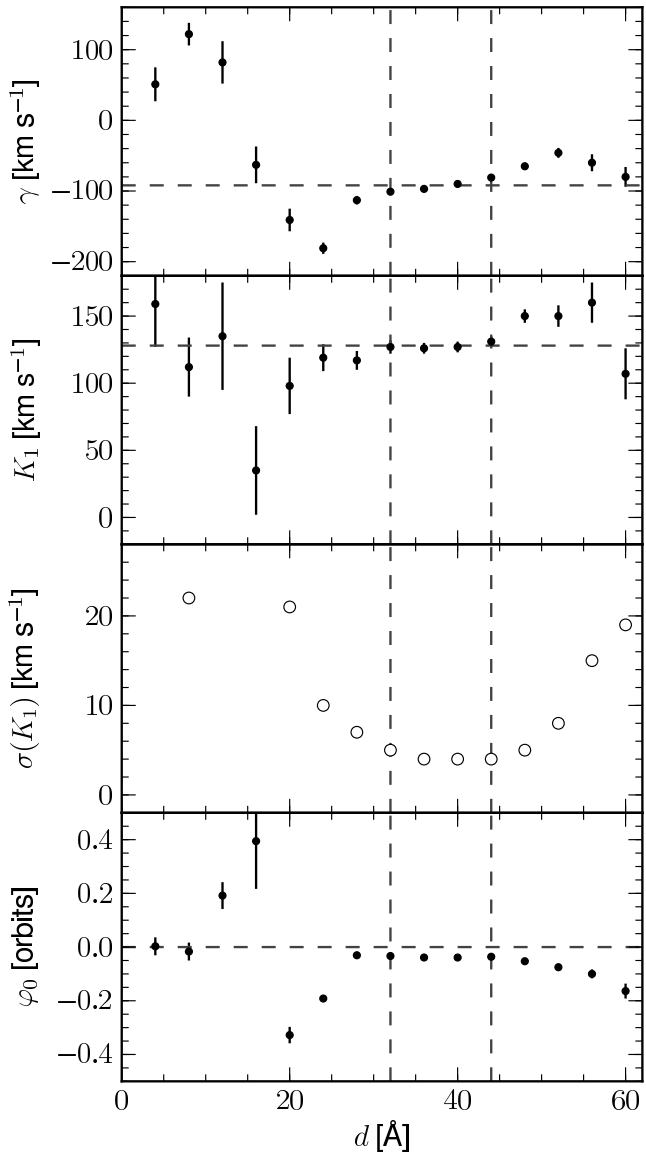}
\caption[]{Diagnostic diagram for the March 27 data. The dashed horizontal
lines in the $\gamma$ and $K_1$ plots mark the adopted average values for
the range of separations $d$ = 32 to 44 {\AA} (dashed vertical lines). In
the plot of the orbital phase (bottom) it instead marks the zero point as
determined by Eq.~(\ref{ephem_eq}).
}
\label{dd_fig}
\end{figure}

\begin{table}
\caption[]{Possible choices for the radial velocity parameters.}
\label{rvpar_tab}
\begin{tabular}{lllll}
\hline\noalign{\smallskip}
$\gamma$      & $K_1$         & $\varphi_0$   & $d$     & Criterion \\%
\hspace{0cm} (km s$^{-1}$) & (km s$^{-1}$) & (orbits)      & ({\AA}) & \\
\hline\noalign{\smallskip}
$-$113(6)     & 117(7)        & $-$0.031(11)  & 28      & $\varphi_0$ \\
$-$92(9)      & 128(3)        & $-$0.037(03)  & 32--44  & $K_1 \sim$ constant\\
$-$46(7)      & 150(8)        & $-$0.075(11)  & 52      & Noise \\
\hline\noalign{\smallskip}
\end{tabular}
\end{table}

\begin{figure}
\includegraphics[width=\columnwidth]{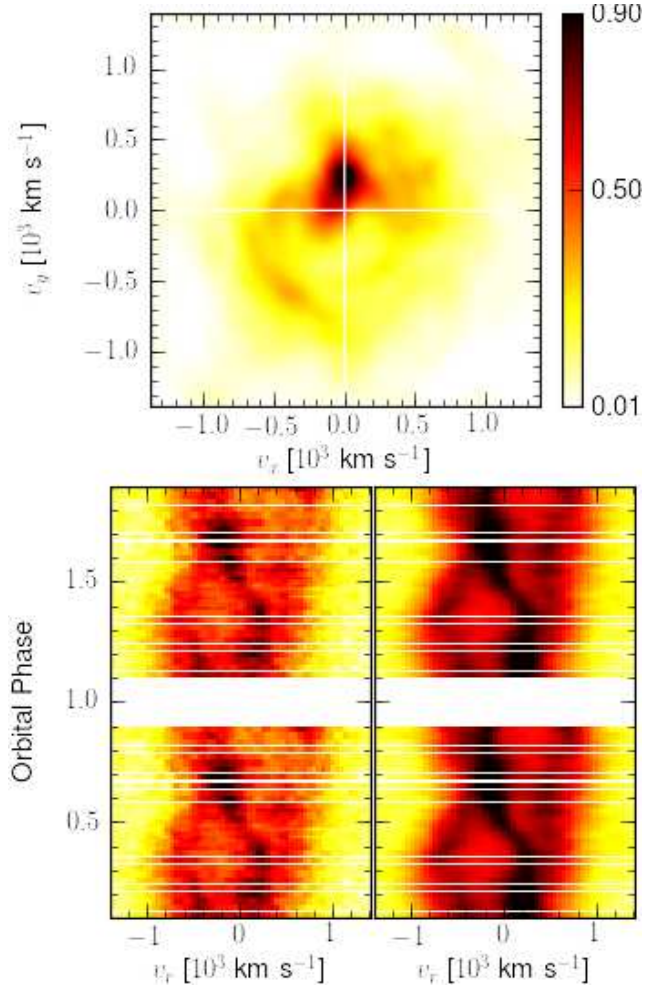}
\caption[]{Doppler map of the H$\alpha$ emission line of the March 27 data
set. The lower plots present the original (left) and the reconstructed (right)
trailed spectra. The intensity is scaled to show the range of 1\% to 90\% of 
the maximum value.}
\label{Hadopn2_fig}
\end{figure}

The H$\alpha$ emission line in our data presents a broad and double-peaked
line profile. We thus measure the radial velocities using
the double-Gaussian technique \citep{schneider+young80-1,shafter83-1}.
Here, the line profile is convolved with two identical Gaussian curves, whose
centres are separated along the wavelength axis by a quantity $d$. The 
wavelength where the fluxes under the two curves are identical yields the
Doppler shift of the line and thus its radial velocity. By varying $d$
and by employing sufficiently narrow Gaussian functions, different parts
of the emission line can be measured. The usual assumption is that the
centre of the emission line represents the part most likely affected by 
potentially present isolated emission sources (e.g., the hot spot or 
the secondary star), while the line wings are mostly formed in the inner disc 
and hence more faithfully reflect the motion of the primary star
\citep[see ][]{horne+marsh86-1}. 

The useful lower limit for the FWHM of the two Gaussians is given by the
spectral resolution of the data. For the grism \#20 set we therefore
use Gaussians with FWHM = 4 {\AA}. As mentioned above, our two data sets
were taken on two subsequent nights during decline from outburst. While the 
system's brightness during that time changed by only slightly more than 0.1 
mag, a visual inspection (Fig.~\ref{tracespecs_fig}) reveals 
significant differences both in the strength of the line and in the line
profile. This is corroborated by comparing the equivalent widths and radial
velocities for corresponding orbital phases. We are thus not allowed
to mix the two data sets, and since the March 26 data do not cover a complete
orbit, we choose the March 27 data for the further analysis, excluding the
spectra around the eclipse (orbital phases 0.9--1.1).

The radial velocities measured in the way outlined above are fitted with a 
sinusoidal function of the form
\begin{equation}
v_r (\varphi) = \gamma - K_1 \sin (\varphi-\varphi_0),
\label{rv_eq}
\end{equation}
where $\gamma$ is the constant term, $K_1$ the semi-amplitude and $\varphi$
the orbital phase with respect to $\varphi_0$. The phases $\varphi$ have
been calculated according to equation (\ref{ephem_eq}), and so one would
expect $\varphi_0 = 0$ for equation (\ref{rv_eq}) describing the motion of the
white dwarf with $\varphi_0$ marking the point of its superior conjunction 
and thus coinciding with the photometric eclipse. In order to identify the 
region of the line that follows the motion of the white dwarf, 
\citet{shafter83-1} introduced the so-called diagnostic diagram, where
the parameters of equation (\ref{rv_eq}) are plotted as a function of the 
separation
$d$ of the two Gaussians. The diagram for our data is shown in 
Fig.~\ref{dd_fig}. The parameters present strong fluctuations at small
separations corresponding to the centre of the line, remain comparatively
constant at medium separations and become variable again at larger 
separations when the noise begins to dominate the line wings. The plot
gives us three possibilities to choose the parameters, summarized in Table
\ref{rvpar_tab}. First, we can use
the point where $\varphi_0$ is closest to 0, corresponding to photometric
eclipse. Ignoring the disturbed central part of the line this is the case
for a separation $d = 28~\mathrm{\AA}$. Secondly, we can take the average for 
the range $d = 32\!-\!44~\mathrm{\AA}$, where $K_1$ is constant. And thirdly, 
$d = 52~\mathrm{\AA}$ represents the last point before the error of the
semi-amplitude $\sigma(K_1)$ increases sharply, thus indicating the start
of the noise-dominated parts. Since a closer inspection shows that Gaussians
with $d = 28~\mathrm{\AA}$ still sample part of the peaks of the line, and 
since the radial velocity curve corresponding to $d = 52~\mathrm{\AA}$ already 
presents a large amount of noise, we favour the second criterion and adopt the
respective parameters for the further analysis.  

Fig.~\ref{Hadopn2_fig} shows the Doppler map for the same data set using
the code from \citet{spruit98-1} with the {\sc idl} environment being replaced
with a {\sc eso-midas} routine \citep{tappertetal03-2}. The input data consist
of the phase-sorted spectra (again excluding the phases around
eclipse) and the $\gamma$ velocity. The latter is iteratively adjusted by
comparing the average input spectrum with the reconstructed one. The final
result was $\gamma = -107~\mathrm{km~s^{-1}}$. Doppler mapping
visualizes the emission distribution in the system in velocity space.
The orientation of the map is such that it shows the system at orbital
phase 0.25, so that the primary star has velocities $(v_x,v_y) = 
(0,-K_\mathrm{WD})$ and the secondary star accordingly $(0,K_\mathrm{RD})$,
where $K_\mathrm{WD}$ and $K_\mathrm{RD}$ are the semi-amplitudes of the
radial velocity curves of the white dwarf and the red dwarf, respectively.
The centre-of-mass is situated at $(0,0)$ and is marked in the plot as the
point of intersection of the vertical and the horizontal line. The two
lower plots show the trailed spectrogram (in phase) from the input data (left)
and the one reconstructed from the Doppler map (right). We see that the
velocity behaviour of the emission components is well reproduced; however,
the variation in intensity is not. This is due to Doppler tomography
assuming that all emission sources are constant in intensity at all phases
\citep{marsh+horne88-1}. This formal prerequisite is in practice almost never 
fulfilled, and especially not in high-inclination systems, where the 
individual emission sources at certain phases are obscured by other 
components. However, in the Doppler map this should only affect the relative 
intensities of the emission sources but not their positions.

We then find a prominent isolated component at velocities ($v_x,v_y$) =
(-25,191) km s$^{-1}$. Since the resolution of the map is 50 km s$^{-1}$ per
pixel, the offset in $v_x$ is sufficiently small to identify the orbital phase
of the emission with that of the secondary star. For a more precise measurement
we have fitted the component with a Gaussian in those spectra in which it was
clearly visible. A Fourier fit to these data yielded 
$K_\mathrm{add} = 221(7)~\mathrm{km~s^{-1}}$ and $\varphi_0 = 0.52(1)$ orbits,
which confirms above interpretation. Apart from the emission lines our grism 
\#16 spectra do not show any evidence of spectral features from the secondary 
out of eclipse, and it thus can be assumed that the observed emission 
component is not due to magnetic activity, but caused by illumination from the 
white dwarf or the accretion disc. The emitting source will then be limited to 
the side of the secondary star facing the white dwarf, and its velocity will 
not correspond to the centre-of-mass of the secondary star, but to a point in 
between that and the Lagrangian Point $L_1$. We will come back to that in 
Section \ref{syspar_sec}.

The area of the isolated emission is somewhat extended towards lower $v_y$ and
more negative $v_x$ velocities, and is likely to represent emission from the 
gas stream or close to the hot spot. The latter itself is optically thick 
and does not present any (significant) contribution to the H$\alpha$ line. 
This can also be seen in the behaviour of the H$\alpha$ flux (lower plot of 
Fig.~\ref{sketch_fig}) which is declining as soon as the hot spot comes into 
the line of sight. We also see that this flux forms a second local minimum
at orbital phase 0.45. A possible explanation is that part of the H$\alpha$ 
emission is produced in the gas stream from the secondary and/or a shock front 
close to the hot spot, and that this part is eclipsed by the hot inner disc 
or the optically thick hot spot (see the sketch in the top panel of 
Fig.~\ref{sketch_fig}).
Emission from the accretion disc is present, although its 
shape is not that of a closed circle but is rather more reminiscent of that 
of spiral waves \citep[e.g., ][]{steeghs01-1}. This would not be an unexpected 
feature in an outbursting CV, but here might just be an artifact, since the 
present data are very limited in both time- and spectral resolution, as well 
as in S/N.

\subsection{System parameters from the spectroscopy}
\label{syspar_sec}

\begin{figure}
\includegraphics[width=\columnwidth]{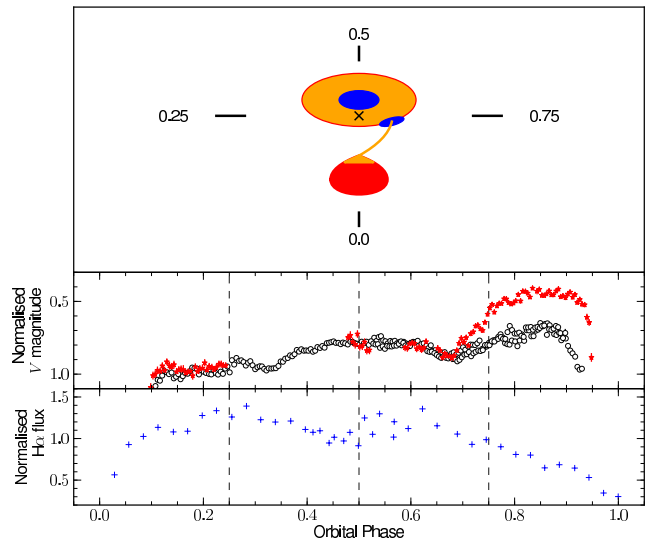}
\caption[]{Top: Sketch of the principal elements of V728 Sco. The orientation
of the system with respect to the observer is indicated for orbital phases
0.0, 0.25, 0.5, and 0.75. Middle: The out-of-eclipse light curves for
2012 March 28 (circles) and April 1 (stars). Both curves have been displaced
vertically to fit into the plot. Bottom: Variation of the flux under the 
H$\alpha$ line throughout the orbit.}
\label{sketch_fig}
\end{figure}

In order to obtain a more detailed picture of V728 Sco we need to know its
system parameters. Since the present data lack completeness and quality
we have to make several assumptions in the process. Consequently, this section 
will not describe the definite but a possible configuration of V728 Sco. Our 
assumptions are the following:
\begin{enumerate}
\item The $K_1$ value extracted from the diagnostic diagram reflects the
motion of the white dwarf. While the double Gaussians did not sample the
extreme wings, the constancy of $K_1$ over a wide range of separations
makes it a good candidate for $K_\mathrm{WD}$.
\item The additional emission component corresponds to an illuminated surface
on the secondary star. The middle plot of Fig.~\ref{sketch_fig} shows a 
close-up of the photometric light curves to emphasize the behaviour 
out of eclipse. The signature of the illuminated surface of the secondary star 
can be clearly seen as a broad maximum opposite the eclipse phase. 
While we do not know the extent of that surface,
it is safe to assume that its velocity $K_2$ lies in between the velocity
of the first Lagrange point $K_\mathrm{L_1}$ and the velocity of the 
centre-of-mass of the red dwarf $K_\mathrm{RD}$, and it probably will be 
closer to the former of those two. At the same time, the secondary star is
elongated due to the Roche lobe deformation, and in the orbital plane the
distance from L$_1$ to the secondary's centre-of-mass is larger than the
equivalent radius of the red dwarf. For the sake of simplicity, we will
assume that the distance $a_\mathrm{RD}$ from the centre-of-mass of 
the secondary to the centre-of-mass of the system is the sum of the distance
of the centre of the additional emission component to the latter and the
radius of the secondary, i.e.~$a_\mathrm{RD} \sim a_\mathrm{add} + 
R_\mathrm{RD}$.
\item The system is seen at a high orbital inclination. This is less an
assumption rather than a conclusion from the presence of a deep and
structured photometric eclipse. We have performed
the subsequent calculations for several inclinations $i \ge 75^\circ$, 
finding only small differences in the respective results. Taking into account
the photometric results (Section \ref{ecl_sec}) we adopt $i = 82^\circ$.
\item Our final assumption regards the type, size and mass of the secondary
star. It is the least certain of all, because it is mainly based on
statistics. \citet{beuermannetal98-1} find, based on the catalogue of
\citet{ritter+kolb03-1}, that the secondaries of the CVs in the period range of
3--4 h have spectral types M3--M5, with most of them being clustered at
M4V. \citet{boyajianetal12-1} conducted an interferometric survey on late-type
dwarfs. Their result for spectral type M4 is based on a single object,
GJ 699, for which they determine $R = 0.19~\mathrm{Ṛ_\odot}$ and $M = 
0.15~\mathrm{M_\odot}$.
In CVs, the respective parameters are likely to be different. A linear-weighted
fit to the secondary masses collected in the \citet{ritter+kolb03-1} 
catalogue (version 3.18) for CVs with orbital periods between 3 and 4~h yields 
the formal relation 
\begin{equation}
M_\mathrm{RD} = -0.125(8) + 3.03(51) P_\mathrm{orb}
\label{mrd_eq}
\end{equation}
for $M_\mathrm{RD}$ in M$_\odot$ and $P_\mathrm{orb}$ in d. For the fit we
have excluded masses without error estimates, as well as three 'outliers'
(UU Aqr, CN Ori, IP Peg). For the orbital period of V728 Sco this yields
$M_\mathrm{RD} = 0.29(5)~\mathrm{M_\odot}$. Simply computing the average for 
the three CVs in the immediate period neighbourhood (DW UMa, V603 Aql, V1315 
Aql, with orbital periods between 3.27 and 3.36 h) yields $M_\mathrm{RD} = 
0.297(6)~\mathrm{M_\odot}$, proving that above fit is reasonable. We then can 
calculate
the radius using the equation (16) from \citet*{kniggeetal11-1} that 
corresponds to our mass and period regime,
\begin{equation}
R_\mathrm{RD} = 0.293(10)~(\frac{M_\mathrm{RD}}{0.2 
\mathrm{M_\odot}})^{0.69(3)}~,
\label{rrd_eq}
\end{equation}
obtaining $R_\mathrm{RD} = 0.38(14)~\mathrm{R_\odot}$.
\end{enumerate}
With these assumptions, we can calculate the velocity of the white dwarf
\begin{equation}
v_\mathrm{WD} = K_1 / \sin i
\label{vwd_eq}
\end{equation}
and its distance to the centre-of-mass 
\begin{equation}
a_\mathrm{WD} = \frac{P_\mathrm{orb} v_\mathrm{WD}}{2 \pi}~.
\label{awd_eq}
\end{equation}
The corresponding distance for the secondary star is 
\begin{equation}
a_\mathrm{RD} = a_\mathrm{add} + R_\mathrm{RD} = 
\frac{P_\mathrm{orb} K_\mathrm{add}}{2 \pi \sin i} + R_\mathrm{RD}~,
\label{ard_eq}
\end{equation}
yielding the distance between the stellar components to
\begin{equation}
a = a_\mathrm{WD} + a_\mathrm{RD}~.
\label{a_eq}
\end{equation}
The velocity of the red dwarf calculates as
\begin{equation}
v_\mathrm{RD} = 2 \pi a_\mathrm{RD} / P_\mathrm{orb}~,
\label{vrc_eq}
\end{equation}
and the mass ratio results to
\begin{equation}
q = v_\mathrm{WD}/v_\mathrm{RD},
\label{q_eq}
\end{equation}
which implies a white dwarf mass of
\begin{equation}
M_\mathrm{WD} = M_\mathrm{RD}/q~.
\label{mwd_eq}
\end{equation}

All these parameters are only weakly dependent on the inclination $i$. The
parameters (\ref{vwd_eq}) and (\ref{awd_eq}) can directly be derived from the 
observed quantities, yielding the ranges
$128.5 < v_\mathrm{WD} < 132.5~\mathrm{km~s^{-1}}$ and $0.35 < a_\mathrm{WD}
< 0.36~\mathrm{R_\odot}$ for $85^\circ > i > 75^\circ$. 
All the remaining parameters depend on the adopted value of
$M_\mathrm{RD}$ and the respective $R_\mathrm{RD}$ from Eq.~(\ref{rrd_eq}).
We determine in (\ref{q_eq}) a mass ratio $q = 0.36$, and according to
(\ref{mwd_eq}) a white dwarf mass $M_\mathrm{WD} = 0.81~\mathrm{M_\odot}$, a 
value
very typical for other CVs. We will adopt these values for the further 
analysis of the system parameters.

\subsection{Analysis of the low-state eclipse}
\label{ecl_sec}

\begin{figure}
\includegraphics[width=\columnwidth]{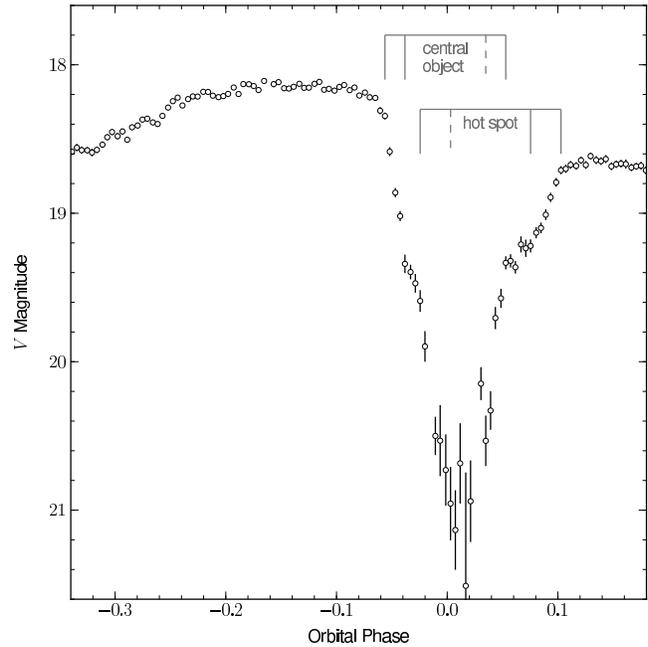}
\caption[]{The ingress and egress times of the central object and the hot spot
as seen in the low-state eclipse. The times indicated by the dashed lines
are not measured, but implied by the width of the corresponding opposite
transition.}
\label{lowecl_fig}
\end{figure}

\begin{table}
\caption[]{Parameters from on the eclipse analysis according to 
\citet{ritter80-3} and the Roche lobe geometry for different orbital 
inclinations $i$.}
\label{phpar_tab}
\begin{tabular}{llllll}
\hline\noalign{\smallskip}
$i$ & $R_\mathrm{RD}/a$ & $a_s/a$ & $R_d/a$ & $\alpha$ & $q$ \\%
\hspace{0cm} ($^\circ$) &            &         &         & ($^\circ$) & \\
\hline\noalign{\smallskip}
90 & 0.256 & 0.819 & 0.325 & 48 & 0.21 \\
85 & 0.271 & 0.834 & 0.318 & 50 & 0.25 \\
82 & 0.291 & 0.852 & 0.312 & 54 & 0.36 \\
80 & 0.309 & 0.866 & 0.308 & 56 & 0.45 \\
75 & 0.364 & 0.898 & 0.300 & 62 & 0.83 \\
\hline\noalign{\smallskip}
\end{tabular}
\end{table}

\begin{table}
\caption[]{Physical parameter estimates for V728 Sco.}
\label{syspar_tab}
\begin{tabular}{ll}
\hline\noalign{\smallskip}
$i$ & 82$^\circ$ \\
$q$ & 0.36 \\
$M_\mathrm{WD}$ & 0.81 M$_\odot$ \\
$M_\mathrm{RD}$ & 0.29 M$_\odot$ \\
$R_\mathrm{RD}$ & 0.38 R$_\odot$ \\
$a$ & 1.31 R$_\odot$ \\
$a_s$ & 1.12 R$_\odot$ \\
$R_d$ & 0.41 R$_\odot$ \\
$R_\mathrm{CO}$ & 0.09 R$_\odot$ \\
$R_\mathrm{HS}$ & 0.10 R$_\odot$ \\
\hline\noalign{\smallskip}
\end{tabular}
\end{table}

As shown in Fig.~\ref{v728scolc_fig}, we observed V728 Sco on 2012 April 1
in a brightness state that is about 1.5 mag fainter than the brightest 
observed value for the post-nova (from 2012 March 25). There are also 
remarkable differences in the corresponding light curve, showing a pronounced 
hump maximum at phase 0.85 and an eclipse light curve with a clear distinction 
of two totally eclipsed, but phase-shifted bodies: first that of a luminous 
source around the white dwarf location, which we will call 'central object'
(suffix CO), and later that of the hot spot (HS) (Fig.~\ref{lowecl_fig}).
Unfortunately, due to the very faint flux at eclipse minimum ($V \sim 
21~\mathrm{mag}$) some of the data points around these phases have rather large 
errors. In spite of this, we here present an attempt to determine some of the 
binary parameters in a preliminary manner.  

For this purpose, we used the geometrical method described in detail by 
\citet{ritter80-3} for the analysis of the data on Z Cha, a dwarf nova with 
similar eclipse characteristics. From our light curve, we derived for V728 Sco 
a total mean duration of 297 s for the ingress and egress of the central 
object, and 378 s for those of the hot spot. The total eclipse 
durations (the differences between the mean ingress and egress epochs) are 974 
s for the central object and 1188 s for the hot spot. Finally, the 
time difference between both mid-eclipses was measured to 568 s. Applying 
equations 8, 9, 10 and 11 of \citet{ritter80-3} we can derive several 
relations between binary parameters for different orbital inclinations, in 
particular the radius of the secondary star $R_\mathrm{RD}$, the binary
separation $a$, the distance between the secondary star and the hot spot
$a_s$, the disc radius $R_d$ as the distance between the white dwarf and 
the hot spot, and the angle $\alpha$ between the hot spot and the secondary
star and with the compact object as vertex. If we adopt that $R_\mathrm{RD}$ 
is equal to the volume radius of the Roche lobe filling secondary, we can 
apply equations (2.5b) and (2.5c) in \citet{warner95-1} in order to derive the 
mass ratio $q = M_\mathrm{RD}/M_\mathrm{WD}$ from $R_\mathrm{RD}/a$. The 
results are listed in Table \ref{phpar_tab}. Apparently, the mass ratio $q$ is 
strongly dependent on $i$ in the present analysis. On the other hand, there is 
only a weak dependence between $q$ and $i$ in our spectroscopic analysis
(Section \ref{syspar_sec}). We therefore use the spectroscopically determined 
mass ratio $q = 0.36$, in order to find the most probable orbital inclination
$i$ for which this mass ratio is valid, according to the eclipse analysis. 
This yields $i = 82^\circ$.

We can now perform an approximate check on the consistency of our results.
Using $R_\mathrm{RD} = 0.38~\mathrm{R_\odot}$ from Eq.~(\ref{rrd_eq}), we 
derive
$a = 1.31~\mathrm{R_\odot}$ from the photometry. This fits well with the 
spectroscopically determined $a = 1.34~\mathrm{R_\odot}$. One has to keep in 
mind,
however, that those results are not completely independent, because they
both depend on our choice for $R_\mathrm{RD}$. Another test can be done
by using $M_\mathrm{RD}$, $M_\mathrm{WD}$, and $P_\mathrm{orb}$ to calculate 
$a$ from Kepler's Third Law,
\begin{equation}
a = \sqrt[3]{\frac{G(M_\mathrm{WD}+M_\mathrm{RD})P_\mathrm{orb}^2}{4 \pi^2}}~,
\label{kepler3_eq}
\end{equation}
where $G$ is the gravitational constant. This results in $a = 
1.16~\mathrm{R_\odot}$. 
This differs from the photometric and spectroscopic results by 11 and 13 per 
cent, 
respectively, which still appears acceptable, regarding the uncertainties 
involved in our calculations of $a_\mathrm{WD}$ and $a_\mathrm{RD}$.

In Table \ref{syspar_tab} we present a first parameter estimation of V728 Sco,
using the photometrically determined value for the binary separation $a$. The 
radii of the central object and the hot spot ($R_\mathrm{CO}$ and
$R_\mathrm{HS}$, respectively) were calculated using equations 16 and 17 
from \citet{ritter80-3}. The radius $R_\mathrm{CO}$ is at least a factor
6 larger than a typical white dwarf radius, an important observational fact 
which will be discussed in Section \ref{disc_sec}. The observed hump maximum 
at phase 0.85 mentioned above would correspond to an angle $\alpha = 54^\circ$,
in agreement with the calculated value in Table \ref{phpar_tab}. This can
be considered as an additional argument in favour of an orbital inclination near
82$^\circ$. Finally, we can compare the spectroscopically determined distance 
from the white dwarf to the centre-of-mass $a_\mathrm{WD} = 
0.36~\mathrm{R_\odot}$ with 
the photometrically derived radius of the accretion disc $R_d = 
0.41~\mathrm{R_\odot}$.
This yields the distance from the hot spot to the centre-of-mass to 
$a_\mathrm{HS} \sim 0.05~\mathrm{R_\odot}$, which corresponds to a 
comparatively small 
Keplerian velocity $v_\mathrm{HS} = 18~\mathrm{km~s^{-1}}$. This fits quite 
well with our interpretation of the elongated extension of the emission from 
the secondary star seen in the $-v_x,+v_y$ quadrant of the Doppler map 
(Fig.~\ref{Hadopn2_fig}). Note that while the latter was taken during decline
from outburst, the photometric analysis corresponds to the low state, and
thus the disc radii can be expected to be different \citep{smak84-1}. Still, 
due to the apparent short outburst recurrence time (Fig.~\ref{v728scoltlc_fig})
the changes in the disc radius during an outburst cycle are probably smaller
than in other CVs.

Since there is only a single eclipse light curve available for our analysis, 
it was not possible to give any error bar on the estimates presented here.
Furthermore it is worth repeating at this point that several parameters
depend strongly on the mass of the secondary star. Its assumed value is
exclusively based on statistics. The parameters in Table \ref{syspar_tab}
should therefore not be taken at face value, and will have to be tested
by future observations.

\section{Discussion}
\label{disc_sec}

It is of special interest that V728\,Sco shows strong evidence of having a
comparatively low mass-transfer rate. Its orbital period $P̣_\mathrm{orb}
= 3.32~\mathrm{h}$ places it right into the regime of the SW Sex stars. 
These CVs are a group of nova-likes that dominate the orbital period range 
2.8--4 h \citep{rodriguez-giletal07-2}. The typical defining SW Sex 
features suggest that these are high mass transfer systems 
\citep*{rodriguez-giletal07-1}. This is supported by their white-dwarf 
temperatures which is an indication for the accretion rate and which exceeds 
the values expected from angular momentum loss through the standard magnetic 
braking scenario \citep{townsley+gaensicke09-1}. Recent investigations have
shown that most of the non-magnetic CVs in this range and in fact all 
nova-likes seem to be of SW Sex type. This includes, e.g., the old nova
RR Pic \citep{schmidtobreicketal03-4}. It has been suggested that 
SW Sex stars represent a stage in the secular evolution of CVs and that,
in general, CVs reaching that period range share their characteristics
(Schmidtobreick et al., in preparation).

In addition, most old novae are indeed high mass-transfer systems 
\citep*{ibenetal92-1}, and so V728 Sco appears as something of an oddball.
However, it is not alone. The old nova XX Tau also shows the spectral
appearance of a low mass-transfer system \citep{schmidtobreicketal05-2}.
\citet{rodriguez-gil+torres05-1} suggest an orbital period close to 3.26 h,
i.e.~in the immediate vicinity to V728 Sco. Another example is the system
V446 Her, which has a longer orbital period \citep[4.94 h;][]%
{thorstensen+taylor00-1}, and shows 'stunted' dwarf-nova-type outbursts
\citep*{honeycuttetal95-1,honeycuttetal98-2} similar to what is implied by
the long-term light curve of V728 Sco. \citet*{honeycuttetal98-3} give a few 
more examples.

This behaviour of some post-novae has been seen as a strong indication for 
the validity of the hibernation model which predicts changes of the 
mass-transfer rate in the evolution of the pre- and post-nova 
\citep{sharaetal86-1,%
prialnik+shara86-1}. In that model the secondary star suffers an enhanced
irradiation by the hot post-eruption white dwarf, is strongly driven out
of its thermal equilibrium, and as a consequence maintains a much larger
volume than according to its mass. The gradual cooling of the white dwarf
permits the secondary to relax into its equilibrium state, ultimately
losing contact with its Roche lobe, thus terminating the transfer of mass
via L$_1$, and so the post-novae enters into 'hibernation'. Hence, post-novae
immediately after eruption are expected to drive a very high mass-transfer
rate that will gradually decrease with time \citep*{kovetzetal88-1}. Once the
mass-transfer rate has crossed the limit that allows for disc instabilities,
dwarf-nova-type outbursts will occur \citep[e.g.,][for a review]{osaki05-1}.
The low-state behaviour of V728 Sco could be interpreted in this context,
especially because we are here confronted with a 'rather-old' nova, more than 
one century after eruption. Perhaps V728 Sco is just entering its transition 
from a high mass-transfer post-nova stage towards the onset of hibernation.

A different explanation for the observed large spread in mass-transfer rates
for CVs was given by \citet{kingetal95-1}. In their approach, the interplay
between irradiation from the white dwarf and an increase in the Roche lobe
radius during the phases of high mass-transfer causes long-term
($\sim 10^5\!-\!10^6~\mathrm{yr}$) oscillations of the secondary's radius,
inducing corresponding variations of the mass-transfer rate. Here,
the fact that most post-novae seem to have high accretion rates would 
simply be a consequence of nova eruptions having a higher probability to 
occur during phases of high mass-transfer. This model predicts identical
behaviour of the pre- and post-novae (beyond the immediate aftereffects of
the nova eruption). With V446 Her, there is at least one low mass-transfer
CV known that represents a strong support for this picture 
\citep{collazzietal09-1}.

Finally, \citet*{schreiberetal00-2} explain the low amplitudes of the outbursts
in post-novae with the system maintaining a hot ionized inner disc due to 
irradiation by the eruption-heated white dwarf. This limits the area of the 
disc that can contribute to the outburst. As the white dwarf cools down, 
the size of the hot inner disc will gradually shrink, the outburst
amplitude will increase, and the outburst frequency will decrease. The very
similar amplitudes of the variation in pre- and post-eruption V446 Her 
\citep{collazzietal09-1} somewhat contradict that picture. Furthermore, there
is as yet no evidence for a change in outburst amplitude or frequency in that 
post-nova \citep*{honeycuttetal11-1}. On the other hand, our analysis of the 
low state eclipse of V728 Sco yields a radius of the central object
$\sim 0.09~\mathrm{R_\odot}$, at least a factor $\sim$6 larger than any white 
dwarf 
radius within the reasonable white dwarf mass range, and even more so for
our estimated mass of 0.81 M$_\odot$ \citep{provencal98-1}. This represents 
strong evidence for the existence of 
the predicted hot inner disc, which is furthermore supported by the presence 
of the strong He{\sc II} $\lambda$4686 emission line in the spectra which 
disappears during eclipse (Section \ref{eclsp_sec}), and thus can be assumed 
to originate in that small part of the disc. Also, the low outburst amplitude 
of $\sim$1.5 mag and the high outburst frequency fit well within the framework 
of the Schreiber et al.~model. Future long-term monitoring is desirable to
properly investigate the characteristics of the outburst behaviour.

\section{Summary}

We have presented photometric and spectroscopic data on the recently
recovered old nova V728 Sco, 150 yr after its eruption, that reveal it 
as an eclipsing system. Its orbital period $P_\mathrm{orb} = 3.32~\mathrm{h}$ 
places it in the midst of a region in the period distribution of CVs that is 
dominated by high mass-transfer systems. Still, the spectroscopic 
characteristics and the long-term variability that resembles dwarf-nova type 
outburst behaviour point to a comparatively low mass-transfer rate. 

The low-state data are of special interest because its light curve is 
characterized by a total eclipse of two different bodies, that of a 'central 
object' surrounding the location of the white dwarf and, phase shifted, that 
of the hot spot. This behaviour is well known for SU UMa type dwarf novae with
$P_\mathrm{orb} < 2~\mathrm{h}$ as Z Cha \citep{warner74-11}, OY Car
\citep{vogtetal81-1} and HT Cas \citep{patterson81-1}, but, to our knowledge, 
it was never observed in old novae or any other type of cataclysmic variable 
above the period gap ($P_\mathrm{orb} > 3~\mathrm{h}$). An inspection of 
published data of about 30 eclipsing old nova remnants revealed always 
V-shaped eclipse light curves, similar to those of V728 Sco at high state 
(Fig.~\ref{v728scolc_fig}), sometimes with short totality phases near eclipse 
centre. The analysis of that low state eclipse provided strong evidence
for the existence of a hot inner disc, as predicted for post-novae
by \citet{schreiberetal00-2}.

V728 Sco could soon play the role of a corner stone in our understanding of 
post-novae and cataclysmic variables in general. This, however, will only be 
possible if more and better photometric and spectroscopic data become available,
especially during the interesting low state. Unfortunately, this is all but
trivial due to the faintness of this target.

\section*{Acknowledgments}
This research was supported by FONDECYT Regular grant 1120338 (CT and NV).
AE acknowledges support by the Spanish Plan Nacional de Astrononom\'{\i}a y 
Astrof\'{\i}sica under grant AYA2011-29517-C03-01. 

We gratefully acknowledge ample use of the SIMBAD database, 
operated at CDS, Strasbourg, France, and of NASA's Astrophysics Data System 
Bibliographic Services. IRAF is distributed by the National Optical Astronomy 
Observatories. 

\bsp

\label{lastpage}

\end{document}